\documentclass[a4paper,twocolumn,english,showpacs,nofootinbib,nobibnotes,aps,pra,floatfix]{revtex4-1}
\usepackage[utf8]{inputenc}
\usepackage{amsmath}
\usepackage{amsthm}
\usepackage{amssymb}
\usepackage{graphicx}
\usepackage[caption=false]{subfig}
\usepackage{xparse}
\usepackage{xcolor}

\makeatletter

\newcounter{thmc}

\newtheorem{proposition}[thmc]{Proposition}
\newtheorem{definition}[thmc]{Definition}

\usepackage{braket}
\usepackage{dsfont}

\makeatother

\usepackage{babel}

\newcommand{\kommentar}[1]{}

\NewDocumentCommand\opti{smmm>{\SplitList{;}}m} {
\begingroup%
\setlength{\belowdisplayskip}{-0.6\baselineskip}%
\IfBooleanTF{#1}{%
    \begin{alignat*}{2}
        & \underset{#3}{\text{#2}} & & #4 \\
        & \text{subject to~~}
        \ProcessList{#5}{ \insertopticonst }
        & &
    \end{alignat*}%
    }{%
    \begin{alignat}{2}
        & \underset{#3}{\text{#2}} & & #4 \\
        & \text{subject to~~}
        \ProcessList{#5}{ \insertopticonst }
        & & \nonumber
    \end{alignat}%
    }%
\endgroup%
}%
\newcommand\insertopticonst[1]{& & #1\\&}

\begin{document}

\title{A protocol for global multiphase estimation}

\author{Giovanni Chesi}
\affiliation{INFN Sezione di Pavia, Via Agostino Bassi 6, I-27100 Pavia, Italy}

\author{Roberto Rubboli}
\affiliation{Centre for Quantum Technologies, National University of Singapore, Singapore}

\author{Alberto Riccardi}
\affiliation{INFN Sezione di Pavia, Via Agostino Bassi 6, I-27100 Pavia, Italy}

\author{Lorenzo Maccone}
\affiliation{Dipartimento di Fisica, Universit\`{a} degli Studi di Pavia, Via Agostino Bassi 6, I-27100, Pavia, Italy \\
INFN Sezione di Pavia, Via Agostino Bassi 6, I-27100, Pavia, Italy}

\author{Chiara Macchiavello}
\affiliation{Dipartimento di Fisica, Universit\`{a} degli Studi di Pavia, Via Agostino Bassi 6, I-27100, Pavia, Italy \\
INFN Sezione di Pavia, Via Agostino Bassi 6, I-27100, Pavia, Italy}

\date{\today}

\begin{abstract}
Global estimation strategies allow to extract information on a phase or a set of phases without any prior knowledge, which is, instead, required for local estimation strategies. We devise a global multiphase protocol based on Holevo's estimation theory and apply it to the case of digital estimation, i.e. we estimate the phases in terms of the mutual information between them and the corresponding estimators. In the single-phase scenario, the protocol encompasses two specific known optimal strategies. We extend them to the simultaneous estimation of two phases and evaluate their performance. Then, we retrieve the ultimate digital bound on precision when a generic number of phases is simultaneously estimated. We show that in the multiphase strategy there is only a constant quantum advantage with respect to a sequence of independent single-phase estimations. This extends a recent similar result, which settled a controversy on the search for the multiphase enhancement. 
\end{abstract}
\maketitle

\section{Introduction}
Quantities that in quantum mechanics are not described by operators require the search for good estimators revealing the value of the corresponding parameters \cite{braunstein}. The efficiency of an estimation strategy is assessed by the Heisenberg bound \cite{ou,qm}, which provides the ultimate achievable precision. 
\\
In this context, we have a general framework for \textit{local} estimation, where an approximate knowledge of the parameter at least is assumed, through, for instance, preliminary coarse measurements that the local estimation aims to improve \cite{paris,barbieri}. The idea is to study the dispersion of the estimator in the neighborhood of the parameter, typically through the Fisher information and the Cramer-Rao bound. The theory has been successfully specified to the quantum formalism and extended to the multiparameter estimation, where more than one parameter is simultaneously estimated \cite{paris,datta,datta2,datta3}.
\\
The same cannot be said for \textit{global} estimation theories, since generally global models are exactly solved only in special cases \cite{rafal}. In global estimation theories, no a priori knowledge of the parameter is assumed and the goal is to investigate the quality of an estimation procedure defined by a measurement on an input state where the parameter is encoded. One can do this by describing parameter and estimator as phases and averaging a suitable $cost function$ over the whole range.
Much effort has been spent to apply a Bayesian approach to local theories, thus accounting for uncertainty on the knowledge of the parameter \cite{vantrees,rafal}. This approach produced many useful bounds on specific cost functions (typically the mean square error), such as the Ziv-Zakai bounds, originated from Ref.~\cite{ziv}. Still, all these criteria ground on a preliminary knowledge about the parameter to be estimated.
For what concerns genuine global Bayesian estimation theories, the seminal works by Holevo \cite{holevo} defined the framework and the first main results for a restricted but significant class of cost functions in the case of a non-degenerate generator and single-parameter estimation. Then, the case of degenerate generator was considered and accomplished in Ref.~\cite{pe}. A first extension to the multiparameter scenario grounded on Holevo's global theory was developed in Ref.~\cite{multipe}. To our knowledge, only a few efforts followed in this direction. A generalization of the Quantum Phase Estimation Algorithm (QPEA) to the multiphase context has recently been provided in Ref.~\cite{pezze}. A Bayesian covariant estimation was provided in Ref.~\cite{rafal}, where the authors find the POVMs optimizing cost functions related to the quantum state fidelity in the specific case of pure and mixed qubits as probes. Interestingly, some years later the same authors exploited a minimax approach to generalize a local phase estimation to the global covariant scenario \cite{rafal2}. In particular, they retrieved the Heisenberg bound scaling on the sum of the squared errors in the multiphase case and proved that the advantage of simultaneous multiphase estimation with respect to independent single-phase estimations amounts to a constant factor. This result opens a controversy, since it contradicts the multiphase enhancement previously claimed both in local frameworks \cite{datta,datta2,datta3}, where an intrinsic advantage scaling with the number of simultaneously estimated phases was found, and in the global framework of the multiphase QPEA \cite{pezze}. 
\\
Here we devise a protocol for global multiphase estimation with commuting generators based on the extension of Holevo's optimal POVM \cite{holevo} to the multiphase case exploited in Ref.~\cite{multipe}. The protocol is more general than other procedures, such as the ones in Refs.~\cite{multipe,rafal2}, because it encompasses diverse well-known global single-phase estimation strategies. Indeed, the protocol fixes the POVM only, so that, by suitably changing the input state, it outputs the statistics related to the corresponding strategy, defined by the desired cost function. Estimation strategies can be generally differentiated according to the application of the encoding unitaries into \textit{parallel} and \textit{sequential} \cite{qm,dqe}. In parallel configurations, we encode the parameter on a number $N$ of inputs employed all together, while in the case of sequential strategy we use the encoding unitary $N$ times on the same input, possibly coupled to ancillary systems. We show how a suitable choice of the probes can reduce our protocol to known optimal parallel and sequential strategies.
\\
In principle, if the estimator is a continuous variable, our protocol sets no limitations on the cost function to be optimized. If the estimator is discrete, the protocol needs to satisfy some assumptions on the cost function, that are detailed in Appendix~\ref{appA}, in order to keep the Bayesian average of the cost function invariant under discretization. Here we take a discrete estimator and apply the protocol to the mutual information between the parameter $\phi$ and the estimator $\tilde{\phi}$, which can be defined as \cite{cover}
\begin{equation}
    I(\tilde{\phi}:\phi)=H(\tilde{\phi})-H(\tilde{\phi}|\phi) 
\end{equation}
where $H(\tilde{\phi})$ is the entropy of the estimator and $H(\tilde{\phi}|\phi)$ is the conditional entropy of the estimator with respect to the parameter. Every estimation strategy is intrinsically dependent on the figure of merit one wants to optimize, since it is the indicator of the quality of the estimation itself. For a global estimation, the mutual information is a good candidate because it does not depend on the parameter and has a clear physical meaning: it is the number of bits shared by parameter and estimator. An estimation theory based on the mutual information is known as \textit{digital estimation}. Digital quantum estimation for sequential and parallel strategies was developed in the single-parameter case in Ref.~\cite{dqe}. As noted by M.~J.~W. Hall in Ref.~\cite{hall}, it is quite surprising that seldom a quantum metrology problem has been expressed in information-theoretic terms, being the mutual information, differently from the Cramer-Rao bound, deeply and directly connected to the original information theory as introduced by Shannon. Together with H.~M. Wiseman, he found interesting equivalences between the information-theoretic framework for phase estimation and the Fisher-information formalism \cite{wiseman2}. In particular, they successfully exploited the Holevo information bound as an Heisenberg limit on the mutual information and found that the latter is limited from above by the entropy of the generator for the probe state. 
\\
Finally and more importantly, our protocol can be naturally extended to the multiparameter case, thus providing a benchmark for multiphase estimation. We start by investigating the most simple non-trivial case, which is the simultaneous estimation of two phases. We find that double-phase estimation slightly outperforms independent single-phase procedures. Hence, we move to the case of a generic number of phases and assess the ultimate limit on multiphase estimation in information-theoretic terms, i.e. we derive Heisenberg bounds on the mutual information as a consequence of the Holevo information bound, which was exploited in this context also in Refs.~\cite{dqe,hall,wiseman2}. This last result helps to clarify the controversy on the multiphase enhancement mentioned above. On the one hand, we find that it is true that the maximum achievable information grows with the number of simultaneously estimated phases. On the other hand, by inspecting the effective information gain for each estimated phase, one finds out that asymptotically it amounts to a constant factor, as claimed in Ref.~\cite{rafal2}.
\\
The paper is organized as follows. In Section~\ref{2} we review the basic framework of a global estimation theory and how it can be extended to account for degenerate eigenstates of the generator. Then we present our protocol in Section~\ref{3}. In Section~\ref{4} we focus on the single-parameter estimation and retrieve two existing strategies from our protocol. Our analysis in this context shows a trade-off between optimal sequential and parallel-separable strategies which, as far as we know, has not been remarked yet. We move to the double-phase estimation scenario in Section~\ref{5}, where we derive a direct generalization of the optimal sequential and parallel protocols retrieved in the previous Section and compare them. Moreover, we evaluate them with respect to the single-phase estimation case. Finally, in Section~\ref{6} we consider the general case of estimating $k$ phases with $N$ encoding operations and investigate how the ultimate bound on precision depends on these parameters. Then, in Section~\ref{7}, we draw our conclusions.

\section{Global estimation theory with degenerate states} \label{2}
Here we review Holevo's global quantum estimation theory \cite{holevo}. The aim is to find a POVM minimizing a given functional describing the price of the estimation, i.e. how much information is lost in the estimation process. In particular, we focus on the so-called \textit{Bayesian uniform mean deviation} \cite{holevo}, i.e.
\begin{equation} \label{bmd}
    C = \int_0^{2\pi}\frac{d\phi}{2\pi}\int_0^{2\pi}\frac{d\tilde{\phi}}{2\pi}c(\phi,\tilde{\phi})\text{Tr}[\rho_{\phi}\Pi_{\tilde{\phi}}]
\end{equation}
where $\phi$ is the parameter to be estimated, $\tilde{\phi}$ the estimator, $c(\phi,\tilde{\phi})$ a cost function and the conditional probability density of measuring $\tilde{\phi}$ given $\phi$ is
\begin{equation}
    \text{Pr}(\tilde{\phi}|\phi) = \text{Tr}[\rho_{\phi}\Pi_{\tilde{\phi}}]
\end{equation}
with $\rho_{\phi}$ the state where the phase is encoded and $\Pi_{\tilde{\phi}}$ an element of the POVM $\Pi=\{\Pi_{\tilde{\phi}}\}_{\tilde{\phi}}$ describing the measurement.
\\
Holevo's POVM is optimal in the context of \textit{covariant} phase estimation problems, i.e. in the case where both state and the measurement are covariant. In particular, the state $\rho_{\phi_0}$ is transformed by a unitary representation of the phase-shift group $U_{\phi}$ as follows
\begin{equation} \label{cond1}
    U_{\phi}\rho_{\phi_0}U^{\dagger}_{\phi} = \rho_{\phi+\phi_0}. 
\end{equation}
The covariance of the POVM, on the other hand, implies
\begin{equation} \label{cond2}
    \text{Tr}[\rho_{\phi_0}\Pi_{\tilde{\phi}}] = \text{Tr}[\rho_{\phi_0+\phi}\Pi_{\tilde{\phi}+\phi}] \quad \forall\,\phi \in [0,2\pi].
\end{equation}
Hence, note that in a covariant estimation problem the conditional probability density $\text{Pr}(\tilde{\phi}|\phi)$ depends just on the difference $\tilde{\phi}-\phi$ since
\begin{equation}
    \text{Tr}[\rho_{\phi}\Pi_{\tilde{\phi}}] = \text{Tr}[\rho_0\Pi_{\tilde{\phi}-\phi}].
\end{equation}
Furthermore, the optimality of Holevo's POVM holds for a wide but specific class of cost functions, that we will call here \textit{Holevo class}. These functions are periodic, even and, again, depend on the difference $\tilde{\phi}-\phi$ only. In other words, their Fourier expansion is given by
\begin{equation} \label{cond3}
    c(\phi,\tilde{\phi}) = c_0 + \sum_{k=0}^{\infty}c_k\cos{[k(\tilde{\phi}-\phi)]}
\end{equation}
with the following further assumptions on the coefficients: $c_0\geq 0$ and $c_k \leq 0 \, \forall \, k\geq 1$. 
\\
Note that, given this set of conditions, the Bayesian mean deviation defined in Eq.~(\ref{bmd}) reduces to an integral over a single variable. However, we will keep it in its general form~(\ref{bmd}) since in the following we will not need all of the conditions that guarantee the optimality of Holevo's POVM.
\\
Holevo showed in Ref.~\cite{holevo} that, if the estimation problem is covariant and the cost function belongs to the Holevo class, the differential POVM minimizing the cost functional in Eq.~(\ref{bmd}) is
\begin{equation} \label{holevopovm}
    d\mu(\tilde{\phi}) = \frac{d\tilde{\phi}}{2\pi}|e(\tilde{\phi})\rangle\langle e(\tilde{\phi})|
\end{equation}
where
\begin{equation}
    |e(\tilde{\phi})\rangle \equiv \sum_{n=0}^{\infty}e^{in\tilde{\phi}}|n\rangle
\end{equation}
are the Susskind-Glogower vectors and $\{|n\rangle\}_n$ is the set of eigenstates of the generator $H$ of the transformations $U_{\phi}$.
\\
The optimal POVM in Eq.~(\ref{holevopovm}) can be generalized to the case where the generator is degenerate through the \textit{projection method} \cite{pe}, 
We can easily show in which sense the generator can be degenerate and how the method works if we focus on the equatorial qubit state 
\begin{equation} \label{eqstate}
    |\psi_0\rangle \equiv \frac{1}{\sqrt{2}}(|0\rangle+|1\rangle).
\end{equation}
and take $N$ identical copies of it as a probe $|\tilde{\psi}_0\rangle$, i.e. $|\tilde{\psi}_0\rangle\equiv |\psi_0\rangle^{\otimes N}$. In such a case, an encoding operation as $U_{\phi}^{\otimes N}$ with generator $H=|1\rangle\langle 1|$ would provide the same eigenvalues for different states defined by the same number of $|1\rangle$ states.
We can remove the degeneracy if we consider the non-degenerate subspace $\mathcal{H_{||}}$ of $\mathcal{H}$ such that $\mathcal{H}=\mathcal{H_{||}}\otimes\mathcal{H}_{\bot}$ and $\mathcal{H_{||}}$ is spanned by the normalized vectors $|n_0,n_1\rangle_{||}\propto P_{n_0,n_1}|\Psi_0\rangle$, where $n_0$ and $n_1$ are the numbers of zeros and of ones respectively, $|\Psi_0\rangle$ an arbitrary pure initial state and $P_{n_0,n_1}$ a projector onto the degenerate eigenspace of dimension $\binom{N}{n_1}$ 
generated from all the states defined by the tensor product of $n_1$ states $|1\rangle$ and $n_0 = N - n_1$ states $|0\rangle$. Then the vectors $|n_0,n_1\rangle_{||}$ spanning $\mathcal{H_{||}}$ can be defined as
\begin{equation} \label{parstates}
\begin{aligned}
   |n_0,n_1\rangle_{||}&= |N-n,n\rangle_{||} \equiv |n\rangle_{||}\\
   &\equiv \frac{1}{\sqrt{\lambda_n}}\sum_{\nu_1=0}^1...\sum_{\nu_N=0}^1\delta\left(\sum_k\nu_k-n\right)\bigotimes_{k=1}^N|\nu_k\rangle
\end{aligned}
\end{equation}
where $\lambda_n\equiv\binom{N}{n}$ is a symmetrization factor accounting for the multiplicity of the corresponding eigenvalue. Note that here we relabeled $n_1 = n$, i.e. $n$ is the number of ones. The POVM can be chosen in a block diagonal form on $\mathcal{H}$, i.e. $d\mu(\tilde{\phi})=d\mu_{||}(\tilde{\phi}) \bigoplus d\mu_{\bot}(\tilde{\phi})$, where $d\mu_{\bot}(\tilde{\phi})$ can be an arbitrary POVM on $\mathcal{H}_{\bot}$, since, being defined on the states orthogonal to $|\Psi_0\rangle$, does not contribute to $\text{Pr}(\tilde{\phi}|\phi)$. Therefore, the problem is reduced to finding the optimal $d\mu_{||}(\tilde{\phi})$. It was proved in Ref.~\cite{pe} that the form of the POVM is identical to the one in the non-degenerate case with the eigenstates of $H$ replaced by the basis of $\mathcal{H}_{||}$ defined in Eq.~(\ref{parstates}), i.e.
\begin{equation} \label{holevopovm2}
    d\mu_{||}(\tilde{\phi}) = \frac{d\tilde{\phi}}{2\pi}|E(\tilde{\phi})\rangle\langle E(\tilde{\phi})|
\end{equation}
with
\begin{equation} \label{glosuss2}
    |E(\tilde{\phi})\rangle \equiv \sum_{n=0}^{N}e^{in\tilde{\phi}}|n\rangle_{||}.
\end{equation}

\section{A multiphase global estimation protocol} \label{3}

The multiphase global estimation protocol we are introducing here fixes the measurement to the Holevo POVM and optimizes a given figure of merit over the set of input states. The algorithm can be described as follows.
\begin{itemize}
    \item Fix a set of commuting generators $H_j$ defining the unitary representations $U_{\phi_{j}}$ of the phase-shift group with the corresponding set of phases to be estimated $\{\phi_j\}_{j=1}^{k}\equiv\pmb{\phi}$
    \item using the procedure described above, build the Holevo POVM $d\mu_{||}(\pmb{\tilde{\phi}})$ on the non-degenerate set of eigenstates of the generators for the set of estimators $\{\tilde{\phi}_j\}_{j=1}^k\equiv\pmb{\tilde{\phi}}$
    \item choose a figure of merit for the estimation process
    \item optimize the figure of merit over the set of input states $\rho_0\in \mathcal{S}(\mathcal{H}_{||})$, with  $\mathcal{S(\cdot)}$ set of quantum states, through the Bayesian mean deviation, i.e. 
    \begin{equation}
        C = \underset{\rho_0}{\text{min}}\left\{\int d\pmb{\phi}\,\text{Tr}\left[\int\,d\mu_{||}(\pmb{\tilde{\phi}})c(\pmb{\phi},\pmb{\tilde{\phi}})U_{\pmb{\phi}}^{\otimes N}\rho_{0}(U_{\pmb{\phi}}^{\dagger})^{\otimes N}\right]\right\}
    \end{equation}
    where $U_{\pmb{\phi}}\equiv \exp{(i\sum_jH_j\phi_j)}$ and the figure of merit is established by the cost function $c(\pmb{\phi},\pmb{\tilde{\phi}})$.
\end{itemize}
As mentioned above, here our figure of merit is the mutual information $I(\pmb{\tilde{\phi}}:\pmb{\phi})$. The mutual information has the opposite meaning of a cost function since it provides the number of bits of the parameters gained through the estimation procedure. However, it is built by definition on a Bayesian mean deviation, with the so-called \textit{surprise} as a cost function, i.e.
\begin{equation} \label{surprise}
    c(\pmb{\phi},\pmb{\tilde{\phi}}) = -\log_{2}\text{Pr}(\pmb{\tilde{\phi}}|\pmb{\phi}).
\end{equation}
Note that, despite being covariant, in general it does not belong to the Holevo class, since it is generally not periodic nor even. The Bayesian mean deviation whose cost function is the surprise is the conditional entropy $H(\pmb{\tilde{\phi}}|\pmb{\phi})$, which provides the number of bits that we have to pay for the estimation. The mutual information is a \textit{gain function}, i.e. it is the difference between the maximum number of phase bits $H(\pmb{\tilde{\phi}})$ that we can get (fixed by the number of resources $N$ that we exploit for the estimation) and the number of bits lost in the estimation procedure, provided by the conditional entropy.
\\
The minimization of the conditional entropy over the set of input states is not a simple task. However, we can get an idea of the optimal pure state in the single-phase case through a local minimization of the cost function in Eq.~(\ref{surprise}) for $\tilde{\phi}\sim\phi$, i.e. we can look for the state $|\Psi_0\rangle$ such that
\begin{equation} \label{locmin}
c(\phi,\tilde{\phi}\sim\phi) \geq \underset{\Psi_0}{\text{min}}\{-\log_{2}|\langle E(\tilde{\phi})|U_{\phi}|\Psi_0\rangle|^2\}.
\end{equation}
Since the logarithm is a monotonic function, this minimization is equivalent to maximizing the distribution $\text{Pr}(\tilde{\phi}\sim\phi|\phi) = |\langle E(\tilde{\phi})|U_{\phi}|\Psi_0\rangle|^2$, which, via Lagrange multipliers, readily gives as optimal state a uniform superposition of the non-degenerate states $\{|n\rangle_{||}\}_{n=0}^N$, known as \textit{Holland-Burnett state} \cite{holland}, i.e.
\begin{equation} \label{optst}
    |\Psi_0\rangle_{\text{opt}} \equiv \frac{1}{\sqrt{N+1}}\sum_{n=0}^N|n\rangle_{||}.
\end{equation}
\\
As mentioned in the Introduction, on the one hand our protocol based on Holevo's POVM allows to retrieve known global single-phase estimation strategies and, on the other hand, it paves the way for multiparameter global estimation.

\section{Single-phase estimation} \label{4}
\subsection{Theory}
Here we show that, by suitably changing the probe state, our protocol allows to retrieve as specific cases two known optimal single-parameter parallel and sequential estimation strategies and, to this aim, we introduce a notion of equivalence between estimation protocols.
\begin{definition} \label{def1}
    Two estimation protocols $A$ and $B$ are \textit{equivalent} if the output probability distribution of the estimator $\tilde{\phi}$ conditioned on the parameter $\phi$ is the same, i.e. if
    \begin{equation}
    \text{Pr}_{A}(\tilde{\phi}|\phi) = \text{Pr}_{B}(\tilde{\phi}|\phi).
\end{equation}
\end{definition}
In particular, we test our protocol with two different probes. By comparing the resulting conditional probabilities with the ones of known single-parameter estimation protocols, we will establish an equivalence through Definition~(\ref{def1}). Hence, we can appreciate the generality of this approach.
\subsubsection{Preparation}
We will consider first a probe state which is known to achieve the standard quantum limit and, second, a state saturating the Heisenberg bound \cite{dqe}.
\\
The former is a probe which is very commonly exploited in parallel schemes. It is defined by $N$ copies of the equatorial state $|\psi_0\rangle$ in Eq.~(\ref{eqstate}), i.e. the separable state
\begin{equation} \label{parprobe}
    |\Psi_0\rangle^{(1)} \equiv |\psi_0\rangle^{\otimes N} = \frac{1}{2^{N/2}}\sum_{n=0}^N\binom{N}{n}|n\rangle_{||}.
\end{equation}
As for the latter, we take the Holland-Burnett states in Eq.~(\ref{optst}) since we found in the previous Section that they locally optimize the mutual information. Namely,
\begin{equation} \label{seqprobe}
    |\Psi_0\rangle^{(2)} \equiv \frac{1}{\sqrt{N+1}}\sum_{n=0}^N|n\rangle_{||}.
\end{equation}
The Holland-Burnett states are entangled (see Appendix~\ref{appB}), which implies that we need the $N$ probe states to be entangled to achieve the Heisenberg bound.  \\
The phase is encoded through $U_{\phi}^{\otimes N}|\Psi_0\rangle^{(x)}\equiv|\Psi_{\phi}\rangle^{(x)}$ with $x=1,2$ and $U_{\phi}\equiv\exp{(iH\phi)}$ with generator $H=|1\rangle\langle 1|$.
\subsubsection{Measurement}
Our protocol provides an estimation of the parameter $\phi$ through Holevo's POVM. The probability density of the estimator $\tilde{\phi}$ conditioned on the value of the parameter $\phi$ can be retrieved from
\begin{equation}
    \text{Pr}(\tilde{\phi}|\phi)\frac{d\tilde{\phi}}{2\pi} = \text{Tr}[d\mu(\tilde{\phi})_{||}\rho^{(x)}_{\phi}]
\end{equation}
with $\rho^{(x)}_{\phi}=|\Psi_{0}\rangle\langle\Psi_{0}|^{(x)}$ and $x=1,2$.
\\
If the phase is encoded on $\rho_{\phi}^{(1)}$, we find that the distribution of the estimator conditioned on the parameter is
\begin{equation}
    \text{Pr}^{(1)}(\tilde{\phi}|\phi) = \left|\frac{1}{\sqrt{2^N}}\sum_{n=0}^N\sqrt{\binom{N}{n}}e^{i(\tilde{\phi}-\phi)n}\right|^2.
\end{equation}
On the other hand, if we take $\rho^{(2)}$, we find
\begin{equation}
    \text{Pr}^{(2)}(\tilde{\phi}|\phi) = \frac{1}{N+1}\frac{\sin^{2}[(N+1)\pi(\phi-\tilde{\phi})]}{\sin^2[\pi(\phi-\tilde{\phi})]}.
\end{equation}
\\
Since our figure of merit is the mutual information, here we consider the case of a discrete estimator $\tilde{\phi}/2\pi = m/(N+1)$ with $m\in [0,N]$. 
We show in Appendix~\ref{appA} how the discretization of the distributions and of the cost functional works. In particular, we find that the conditional probability density is modified according to the prescription
\begin{equation} \label{presc}
    p(m|\phi) = \frac{1}{N+1}\text{Pr}\left(\tilde{\phi}=\frac{2\pi\,m}{N+1}\bigg\rvert\phi\right)
\end{equation}
with the normalizations $\sum_{m=0}^Np(m|\phi)=1$ and $\int_0^{2\pi}p(\tilde{\phi}|\phi)=1$.

\subsection{Results}
\subsubsection{Equivalence}
By applying the prescription in Eq.~(\ref{presc}), we find that the conditional probability densities related to the initial states $|\Psi_0\rangle^{(1)}$ and $|\Psi_0\rangle^{(2)}$ for a discrete estimator $m$ are
\begin{equation}
\begin{aligned}
    p^{(1)}(m|\phi) = \frac{1}{N+1}\left|\frac{1}{\sqrt{2^N}}\sum_{n=0}^N\sqrt{\binom{N}{n}}e^{i(m/(N+1)-\phi)n}\right|^2
\end{aligned}
\end{equation}
and
\begin{equation}
\begin{aligned}
    p^{(2)}(m|\phi) = \frac{1}{(N+1)^2}\frac{\sin^{2}[(N+1)\pi\phi]}{\sin^2[\pi(\phi-m/(N+1))]}.
\end{aligned}
\end{equation}
We can recognize in the Equations above the probability densities of two well-known estimation protocols. 
\\
The first one, $p^{(1)}(m|\phi)$, is the conditional distribution that describes optimized \textit{parallel-separable strategies} \cite{dqe}.
In parallel strategies the $N$ probes are employed jointly, i.e. the transformation $U_{\phi}$ is applied to all the probes together \cite{qm}. If the
input state is not entangled, we say that the strategy is separable. It is not surprising that we end with the distribution of the optimal parallel-separable strategy since the probe state in Eq.~(\ref{parprobe}) and the Holevo POVM exactly describe that strategy.
Indeed, it is simple to see that the POVM $\{\Pi_{\tilde{\phi}}\}_{\tilde{\phi}}$ with $\Pi_{\tilde{\phi}}\equiv|E(\tilde{\phi})\rangle\langle E(\tilde{\phi})|$, defined after the differential POVM in Eq.~(\ref{holevopovm2}), is a particular case of Davies' POVM \cite{davies,dqe} maximizing the mutual information in parallel-separable strategies. \\
On the other hand, the second distribution is the same as the optimal \textit{sequential strategy}, whose typical application is the well-known QPEA \cite{qm,dqe,mosca,nielsen}. The basic scheme of a sequential strategy is a single probe undergoing sequentially $N$ times the transformation $U_{\phi}$, in the presence of ancillary systems. Note that in this case our protocol is radically different from the standard QPEA in terms of preparation and measurement. However, we can establish an operational equivalence between the two protocols based on Definition~(\ref{def1}), i.e. we can say that running our protocol with the locally optimal state in Eq.~(\ref{seqprobe}) is equivalent to the optimal sequential strategy. 

\subsubsection{Performance}

\begin{figure}
    \centering
    \includegraphics[scale=0.25]{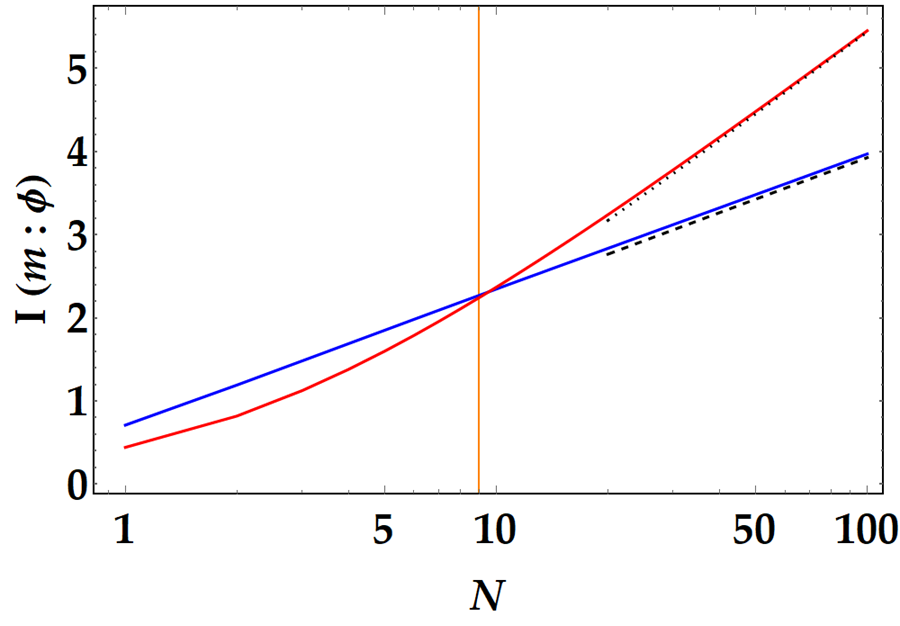}
    \caption{Base-$e$ log-linear plot of the mutual information $I(m:\phi)$ between estimator, $m$, and parameter, $\phi$, in the cases of sequential (red line) and parallel-separable (blue line) strategy. The dotted line displays the asymptotic behavior of the former, the dashed one the asymptotic of the latter.}
    \label{fig:mi2D}
\end{figure}

The performance of sequential and parallel protocols has been largely explored \cite{qm,dqe,pezze, mosca,nielsen}. However, our digital approach can provide a deeper understanding on the topic and a basic reference to be compared to the multiphase case that we will inspect in the next Sections.
Firstly, let us inspect the distribution $p^{(1)}(m|\phi)$.
As far as we know, there is no closed form for that. However, 
we note that for large $N$ a single period of $p^{(1)}(m|\phi)$ can be approximated as a Gaussian distribution, i.e.
\begin{equation}
    p^{(1)}(m|\phi) \xrightarrow[N\gg1]{}\frac{\sqrt{2\pi N}}{N+1}\exp{\left[-2N\pi^2\left(\phi-\frac{m}{N+1}\right)^2\right]}.
\end{equation}
In this case, the mutual information obtained from the tensor-product state $|\Psi_0\rangle^{(1)}$ is asymptotic to the standard quantum limit $\text{SQL}(N) = (\log_2N)/2$ \cite{dqe}. In particular, we find \begin{equation}
\begin{aligned} \label{parbound}
    I^{(1)}(m:\phi)\xrightarrow[N\gg1]{}& \,\text{SQL}(N)+\frac{1}{2}\left(\log_2(2\pi)-\frac{1}{\ln{2}}\right) \\
    \sim & \,\text{SQL}(N) + 0.6044.
\end{aligned}
\end{equation} On the contrary, the mutual information of the QPEA was found to be asymptotically nearly optimal, in the sense that it tends to the Heisenberg bound
$
\text{HB}(N) = \log_2(N+1)
$ \cite{dqe}.
More specifically, the mutual information provided by the QPEA, or, equivalently, by our protocol started with $|\Psi_0\rangle^{(2)}$ as a probe state, displays the behavior \begin{equation}
\begin{aligned} \label{seqbound}
    I^{(2)}(m:\phi)\xrightarrow[N\gg1]{} &\,\text{HB}(N) - 2\left(1-\frac{\gamma+\ln{2}-1}{\ln{2}}\right) \\
    \sim\, & \,\text{HB}(N) - 1.2199
\end{aligned}
\end{equation} where $\gamma$ is the Euler-Mascheroni constant.
\\
These limits are displayed in Fig.~(\ref{fig:mi2D}) as a dotted and dashed black line respectively. We note the well-known asymptotic optimality of the sequential strategies over parallel-separable ones \cite{qm,dqe}, also apparent from the limits in Eqs.~(\ref{parbound}) and~(\ref{seqbound}).
Interestingly, we remark that for small $N$ here we find a trade-off: if $N\leq 9$ the mutual information related to the parallel-separable protocol (blue line) is slightly larger ($<1$ bit) than the one corresponding to the sequential protocol (red line). This trade-off clearly outlines the limits of a local optimization such that the one proposed in Eq.~(\ref{locmin}).

\section{Double-phase estimation} \label{5}
\subsection{Theory}
Holevo's POVM can be straightforwardly extended to account for more than one estimator \cite{multipe}. Hence, we can readily generalize the single-phase analysis developed in the previous Section to the multiphase scenario.
Here we focus on the simplest case, i.e. the estimation of two phases, $\phi$ and $\theta$, through the corresponding discrete estimators $m_{\phi}$ and $m_{\theta}$. Differently from Ref.~\cite{multipe}, we investigate the case of a discrete estimator, as mentioned above, and the figure of merit that best conveys the meaning of digital precision is the mutual information, rather than the fidelity of the estimated state with respect to the probe, which is exploited there.
In addition, we do not restrict the analysis to the case of a tensor-product probe state. Thanks to the optimization procedure performed applying our protocol, we can focus on the input state that locally optimizes our figure of merit.

\subsubsection{Preparation}
Consider the qutrit non-degenerate space $\mathcal{H}_{||}$ spanned by the normalized vectors
$$
|n_0, n_1, n_2\rangle_{||} = |N - n_1 - n_2, n_1, n_2\rangle_{||} \equiv |n_1, n_2\rangle_{||}
$$
in analogy with the projection method devised in the qubit case \cite{multipe}.
Again, we take as inputs on the one hand a tensor-product state and, on the other hand, a uniform superposition of the basis elements $|n_1,n_2\rangle_{||}$. Explicitly,
\begin{equation} \label{st1}
    |\Psi_0\rangle^{(1)} = \frac{1}{\sqrt{3^N}}\sum_{n_1=0}^{N}\sum_{n_2=0}^{N-n_1}\sqrt{\binom{N}{n_1,n_2}}|n_1,n_2\rangle_{||}
\end{equation}
and
\begin{equation} 
\label{entangled}
    |\Psi_0\rangle^{(2)} = \frac{1}{\sqrt{M}}\sum_{n_1=0}^{N}\sum_{n_2=0}^{N-n_1}|n_1,n_2\rangle_{||}
\end{equation}
where $M=\sum_{n_1=0}^N\sum_{n_2=0}^{N-n_1}1=(N+1)(N/2 + 1)$. 
The entanglement of the Holland-Burnett qutrit states $|\Psi_0\rangle^{(2)}$ is discussed in Appendix~\ref{appB}.
The unitary encoding $\phi$ and $\theta$ can be represented in the form $U_{\phi,\theta}=e^{i(\phi H_1+\theta H_2)}$ with generators $H_1=|1\rangle\langle 1|$ and $H_2=|2\rangle\langle 2|$. Given this representation, one applies $U_{\phi,\theta}^{\otimes N}$ to the probe state.
\\
As for the qubit case, we show that the $N$ copies of the equatorial qutrit state $|\Psi_0\rangle^{(1)}$ used as probes for our protocol achieve the standard quantum limits, whereas the Holland-Burnett state $|\Psi_0\rangle^{(2)}$ provides a precision asymptotic to the Heisenberg bound. In other words, our protocol run with the former probe generalizes the single-phase optimal parallel-separable strategy, while, if it is started with the latter probe, it generalizes the single-phase optimal sequential strategy.  As for the qubit case, this shows that, to achieve the Heisenberg bound, we need the $N$ probes to be entangled. In this simple example, the generalization of the single-phase case is limited to two phases, but the protocol is amenable for the application to any number of phases. We will define the ultimate limits attainable if a generic number $k$ of phases is jointly estimated with $N$ resources in the next Section.
\subsubsection{Measurement}
Holevo's POVM in the case of two estimators $\tilde{\phi}$ and $\tilde{\theta}$ is a projection over Susskind-Glogower vectors developed on the elements of the non-degenerate qutrit Hilbert space, i.e.
\begin{equation} \label{holevopovm2}
    d\mu_{||}(\tilde{\phi},\tilde{\theta})=\frac{d\tilde{\phi}}{2\pi}\frac{d\tilde{\theta}}{2\pi}|E(\tilde{\phi},\tilde{\theta})\rangle\langle E(\tilde{\phi},\tilde{\theta})|
\end{equation}
with
\begin{equation}   
    |E(\tilde{\phi},\tilde{\theta})\rangle\equiv\sum_{n_1=0}^N\sum_{n_2=0}^{N-n_1}e^{i(n_1\tilde{\phi}+n_2\tilde{\theta})} |n_1,n_2\rangle.
\end{equation}
The discretization procedure works similarly to the single-parameter case, through the prescription
\begin{equation}
\begin{aligned}
 &p(m_{\phi}, m_{\theta}|,\phi,\theta) = \\
 &=\frac{1}{(N+1)^2}\text{Pr}\left(\tilde{\phi}=\frac{2\pi\,m_{\phi}}{N+1},\tilde{\theta}=\frac{2\pi\,m_{\theta}}{N+1}\bigg\rvert\theta,\phi\right). 
 \end{aligned}
\end{equation}

\subsection{Results}
\subsubsection{Performance}
The input states $|\Psi_0\rangle^{(1)}$ and $|\Psi_0\rangle^{(2)}$ in our protocol provide, respectively, the conditional densities

\begin{equation} \label{multitens1}
\begin{aligned}
&p^{(1)}(m_{\phi},m_{\theta}|\phi,\theta) = \frac{1}{(N+1)^2}\bigg\rvert\frac{1}{\sqrt{3^N}}\sum_{n_1=0}^N\sum_{n_2=0}^{N-n_1}\sqrt{\binom{N}{n_1,n_2}} \\
&\exp{\left(2\pi i\left[n_1\left(\frac{m_{\phi}}{N+1}-\phi\right)+n_2\left(\frac{m_{\theta}}{N+1}-\theta\right)\right]\right)}\bigg\rvert^2
\end{aligned}
\end{equation}
and
\begin{equation} \label{multitens2}
\begin{aligned}
&p^{(2)}(m_{\phi},m_{\theta}|\phi,\theta) = \frac{1}{(N+1)^3(N/2+1)}\bigg\rvert\sum_{n_1=0}^N\sum_{n_2=0}^{N-n_1}\\
&\exp{\left(2\pi i\left[n_1\left(\frac{m_{\phi}}{N+1}-\phi\right)+n_2\left(\frac{m_{\theta}}{N+1}-\theta\right)\right]\right)}\bigg\rvert^2.
\end{aligned}
\end{equation}
The distribution $p^{(1)}$ can be again approximated within a single period by a Gaussian for $N\gg 1$, i.e.
\begin{equation}
\begin{aligned}
    &p^{(1)}(m_{\phi},m_{\theta}|\phi,\theta)\xrightarrow[N\gg1]{}\frac{8\sqrt{3}}{9}\pi\frac{N}{(N+1)^2} \\
    &e^{-\frac{16}{9}N\pi^2\left[\left(\frac{m_{\phi}}{N+1}-\phi\right)^2+\left(\frac{m_{\theta}}{N+1}-\theta\right)^2-\left(\frac{m_{\phi}}{N+1}-\phi\right)\cdot\left(\frac{m_{\theta}}{N+1}-\theta\right)\right]}.
    \end{aligned}
\end{equation}
The sum defining distribution $p^{(2)}$ can be analytically calculated and provides
\begin{equation}
    \begin{aligned}
        &p^{(2)}(m_{\phi},m_{\theta}|\phi,\theta)= \\
        &\frac{1}{M}[S^2_{\Delta\phi}+S^2_{\Delta\theta}+S^2_{\Delta\phi-\Delta\theta}-2S_{\Delta\phi}S_{\Delta\theta}C_{(3+2N)(\Delta\phi-\Delta\theta)} \\
        &+2S_{\Delta\phi-\Delta\theta}(S_{\Delta\theta}S_{(3+2N)\Delta\phi}-S_{\Delta\phi}C_{(3+2N)\Delta\theta})]
    \end{aligned}
\end{equation}
where
\begin{align}
&M=8(N+2)(N+1)^{3}S^2_{\Delta\phi}S^2_{\Delta\theta}S^2_{\Delta\phi-\Delta\theta} \\
&S_x\equiv \sin(\pi x) \\
&C_x \equiv \cos(\pi x) \\
&\Delta\alpha \equiv \alpha - \frac{m_{\alpha}}{N+1} \quad \alpha = \phi,\theta. 
\end{align}
\begin{figure}
    \centering
    \includegraphics[scale=0.27]{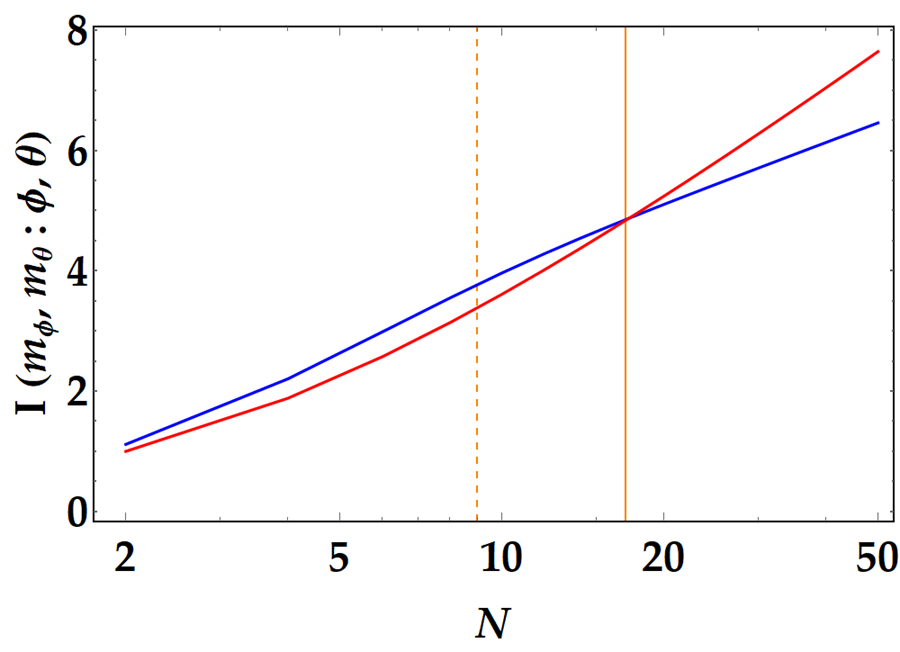}
    \caption{Base-$e$ log-linear plot of the mutual information in the double-parameter estimation case. Blue line: double-phase optimal parallel-separable strategy; red line: double-phase optimal sequential strategy.}
    \label{multicomp}
\end{figure}
\\
Note that the distributions above are obtained by directly generalizing the qubit probe states employed for the single-phase estimation to the qutrit case suited for double-phase estimation. Therefore, we will refer to the estimation strategies related to $p^{(1)}$ and $p^{(2)}$ as the double-phase optimal parallel-separable and sequantial strategy respectively.
\\
We start by comparing the mutual information $I(m_{\phi},m_{\theta}:\phi,\theta)$ provided by $p^{(1)}$ and $p^{(2)}$ for small $N$ ($<20$). The numeric evaluation shown in Fig.~(\ref{multicomp}) reveals that we find again a trade-off between the protocol started with $|\Psi^{(1)}\rangle$ and the one started with $|\Psi^{(2)}\rangle$. In particular, the first slightly outperforms the second up to a threshold on $N$, which in the double-phase case is nearly twice the single-phase threshold.
\\
If we focus on the asymptotics, we find that our protocol run with the tensor-product state $|\Psi^{(1)}\rangle$ is limited by $2\text{SQL}(N)$, whereas if we start it with the Holland-Burnett state $|\Psi^{(2)}\rangle$ we achieve $2\text{HB}(N)$, where $\text{SQL}(N)$ and $\text{HB}(N)$ were defined in the previous Section for the single-phase case. 
Explicitly,
\begin{equation} \label{parbound2}
\begin{aligned}
I^{(1)}(m_{\phi},m_{\theta}:\phi, \theta) \xrightarrow[N\gg1]{}& \,\,2\text{SQL}(N) + \log_2\left(\frac{8\sqrt{3}\pi}{9}\right)-\frac{1}{\ln 2} \\
\sim\, & \,\,2\text{SQL}(N) + 0.8314
\end{aligned}
\end{equation}
and
\begin{equation} \label{seqbound2}
I^{(2)}(m_{\phi},m_{\theta}:\phi, \theta) \xrightarrow[N\gg1]{} \,\,2\text{HB}(N)-3.7899
\end{equation}
where we derived the limit of $I^{(1)}$ by means of an analytic approximation while the limit on $I^{(2)}$ is a numeric result obtained taking $N=500$.
\subsubsection{Comparison with single-phase estimation}
\begin{figure}
    \centering
    {\includegraphics[scale=0.27]{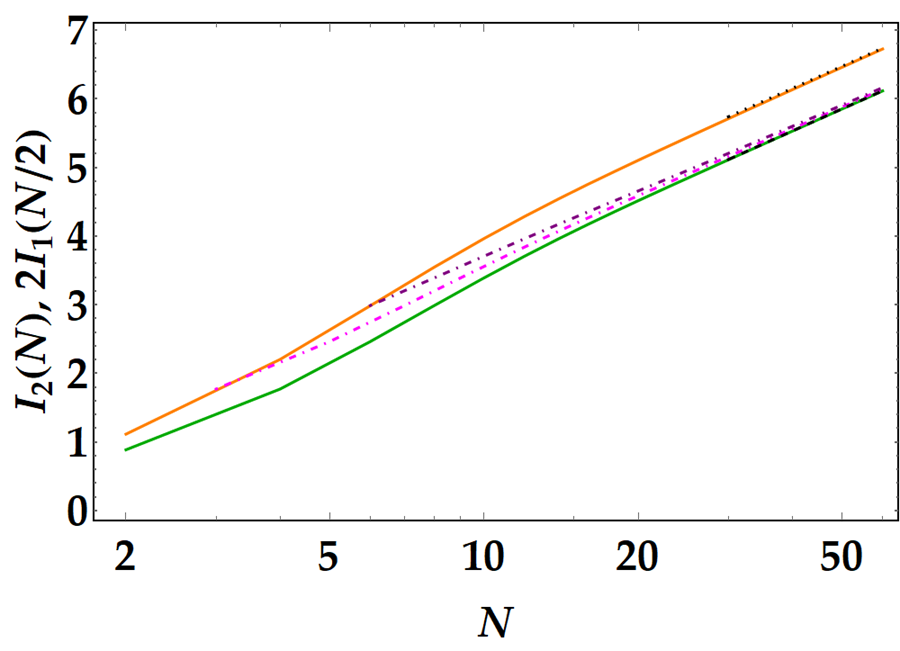}
    \caption{Double-phase optimal parallel-separable strategy. Base-$e$ log-linear plot of the mutual information in the double-phase estimation case $I_2(N)$ (orange line) compared with the mutual information of two independent single-phase estimations $2I_1(N/2)$ where the same number of resources for each phase is employed (green line). The black dotted line displays the asymptotic behavior of the former, while the black dashed line refers to the asymptotic behavior of the latter. We also show $I_1((N+1)/2)$ (magenta dashed-dotted line) and $I_1((N+2)/2)$ (purple dashed-dotted line) and note that, for $N>3$ and for $N>6$ respectively, generalized parallel-separable strategy with double-phase estimation can outperform the corresponding single-parameter estimation even if the latter is supported with one or two probes more for each phase.}
    \label{parallel_double_vs_single}}
    \vspace{0.2cm}
    {\includegraphics[scale=0.27]{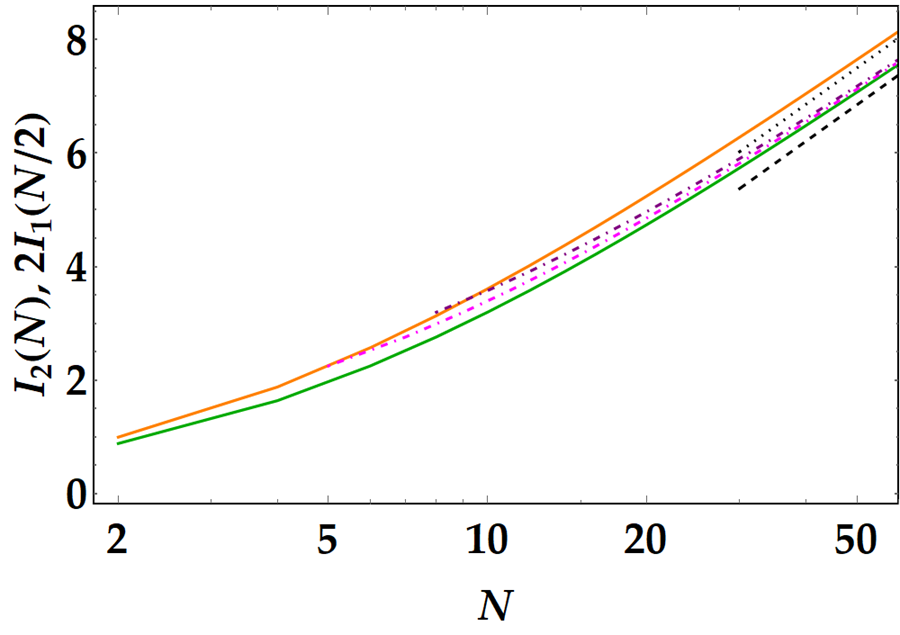}
    \caption{Double-phase optimal sequential strategy. Base-$e$ log-linear plot of the mutual information in the double-parameter estimation case $I_2(N)$ (red line) compared with the mutual information of two independent single-parameter estimations $2I_1(N/2)$ where the same number of probes per parameter were employed (blue line). The black dotted line displays the asymptotic behavior of the former, while the black dashed line refers to the asymptotic behavior of the latter. We also show $I_1((N+1)/2)$ (magenta dotted line) and $I_1((N+2)/2)$ (purple dotted line) and note that, for small $N$, generalized sequential strategy with double-parameter estimation can outperform the corresponding single-parameter estimation even if the latter is supported with one or two probes more.}
    \label{sequential_double_vs_single}}
\end{figure}

Now we investigate if we can find any advantage in estimating two phases simultaneously rather than performing two independent single-phase estimations.
\\
For the sake of clarity, heron we will use for the mutual information between $k$ parameters and estimators the shorthand notation $I_k\equiv I(\pmb{m}:\pmb{\phi})$.
\\
We have to compare double-phase estimation with two single-phase estimation sharing the same total number of encoding operations $N$ to see if there is an advantage in terms of mutual information in estimating two phases simultaneously by encoding them on the same state or in estimating them one by one. Note that if the phases are estimated independently through two distinct single-phase estimation protocols, the mutual information between estimators and phases is simply the sum of the mutual information between each estimator and the corresponding phase, i.e.
\begin{equation}
    I(m_{\phi},m_{\theta}:\phi,\theta)=I(m_{\phi}:\phi)+I(m_{\theta}:\theta) = 2I(m_{\phi}:\phi) \label{2singind}
\end{equation}
where the second equality holds only if the same number of states $N/2$ is employed for the estimation. Shortly, using our new notation, we are comparing $I_2(N)$ with $2I_1(N/2)$.
\\
Figures~(\ref{parallel_double_vs_single}) and~(\ref{sequential_double_vs_single}) show that for small $N$ the simultaneous encoding on qutrits provides a slight advantage over independent single-phase estimations, even if the latter is supported with one, for $N>3$, or two, for $N>6$, more resources for each phase. This result holds for both generalized optimal parallel-separable and sequential strategies. 
The asymptotics shown above prove that exploiting a large number of resources does not improve much the performance of double-phase estimation with respect to two separate single-phase estimations. Indeed, we remark that the advantage does not scale with $N$ and approaches to a constant factor. Specifically, comparing the asymptotics in Eqs.~(\ref{parbound}) and ~(\ref{seqbound}) with the ones in Eqs.~(\ref{parbound2}) and ~(\ref{seqbound2}), we note that the difference in terms of mutual information amounts approximately to one bit. 
\\
One may still wonder if this advantage, which is small in the case of double-phase estimation, depends at least on the number of simultaneously estimated phases. This is the problem of multiphase estimation that we mentioned in the Introduction.
In the next Section, we show that the ultimate bound on the achievable precision in terms of mutual information does depend on the number of jointly estimated phases, but the actual gain with respect to repeated single-phase estimation still amounts to a constant factor at most.

\section{Digital Heisenberg bound on multiphase estimation} \label{6}
\begin{figure}
    \centering
    \includegraphics[scale=0.375]{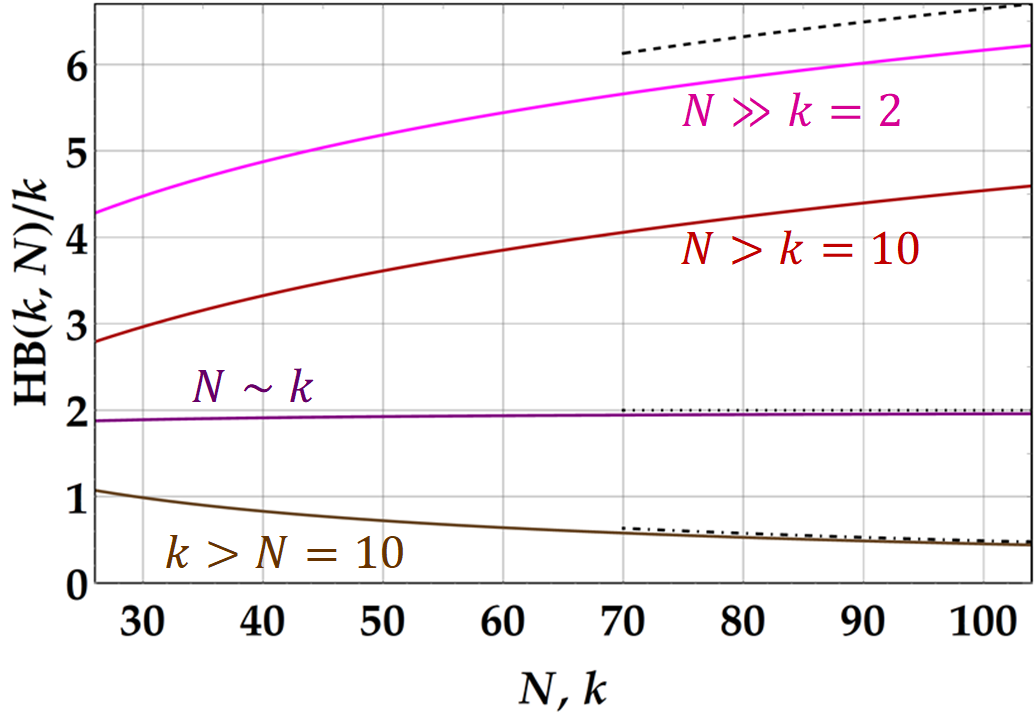}
    \caption{The Heisenberg bound in Eq.~(\ref{hb}) for the simultaneous estimation of $k$ phases with $N$ resources. The bound displays three different scalings according with the ratio $N/k$. First case: $N/k \gg 1$, magenta ($k=2$) and red ($k=10$) lines. The asymptotic behavior of the bound is expressed by Eq.~(\ref{hb1}). The dashed black line displays the ultimate limit $\text{HB}(1,N)$. Second case: $N/k \sim 1$, purple line. The asymptotic behavior of the bound is expressed by Eq.~(\ref{hb2}), i.e. it approaches a plateau emphasized by the dotted black line. Third case: $N/k < 1$, brown line ($N=10$). The asymptotic behavior of the bound is expressed by Eq.~(\ref{hb3}).
    }
    \label{hbfig}
\end{figure}

Finally, we derive the Heisenberg bound on the mutual information $I_k(N)$ between estimators and phases in the case where $N$ resources are employed and $k$ phases simultaneously estimated. Then, we find the number of bits shared between phases and estimators when the $k$ phases are estimated separately, again for fixed number of resources. Hence, we discuss if there actually is an advantage by exploiting multiphase estimation with respect to the straightforward application of a single-phase estimation algorithm for each phase.
\\
We start by stating a Proposition that generalizes for a generic number of simultaneously estimated phases the Heisenberg bound on the mutual information which was derived in Ref.~\cite{dqe} in the case of single-phase estimation. The result is a consequence of the Holevo bound.
\begin{proposition} \label{prop2}
    Given a set of $k$ phases $\pmb{\phi}$ to be estimated with $N$ resources, the ultimate bound on the mutual information $I_k(N)$ between the phases and the corresponding set of estimators $\pmb{\tilde{\phi}}$ reads
    \begin{equation}
        I_k(N) \leq \log_2\binom{N+k}{N} \equiv \text{HB}(k,N) \label{hb}
    \end{equation}
\end{proposition}
The proof can be found in Appendix~\ref{appC}.
\\
From the Heisenberg bound in Eq.~(\ref{hb}) we find that the performance of a multiphase estimation protocol in this context strongly depends on the choice of the parameters $N$ and $k$. 
Depending on the ratio $N/k$, we can devise three specific asymptotic scenarios. We can evaluate the largest mutual information attainable for each jointly estimated phase $\text{HB}(k,N)/k$ for each one. \\
First, if we fix the number of phases and take $N/k\gg 1$, we immediately find for each jointly estimated phase the Heisenberg bound of the single-phase estimation. 
Namely,
\begin{equation}
    \text{HB}(k,N)/k \xrightarrow[N\gg k]{} \,\, \log_2N - \frac{1}{k}\log_2{k!} \sim \text{HB}(1,N). \label{hb1}
\end{equation}
\\
Second, we take $N/k\sim 1$. We observe that this case is quite a typical choice since one may want to use every encoding operation for a different phase and take $k=N$. In such case, we can also explore the asymptotic limit of the Heisenberg bound through Stirling's approximation and find
\begin{equation}
    \text{HB}(k,N)/k \sim \frac{1}{N}[\log_2{(2N)!}-2\log_2{N!}] \xrightarrow[N \gg 1]{} \,\, 2. \label{hb2}
\end{equation}
Finally, we can fix $N$ and take the limit $N/k \ll 1$. We get 
\begin{equation}
    \text{HB}(k,N)/k \xrightarrow[k/N\gg 1]{} \,\, \frac{N}{k}\log_2{\left[e\left(1+\frac{k}{N}\right)\right]}. \label{hb3}
\end{equation}
The collapse of the ratio between the upper bound on the mutual information and the number of phases in Eq.~(\ref{hb3}) is expected since in this case the number of phases is larger than the number of encoding operations, so that there are $k-N$ phases that cannot be distinguished from the others.
\\
These three different regimes are displayed in Fig.~(\ref{hbfig}). We conclude that an efficient multiphase estimation needs to set the number of resources much larger than the number of simultaneously estimated phases.
In this optimal regime, we compare the ultimate bound for multiphase estimation of $k$ phases with $N$ resources in Eq.~(\ref{hb}) with the expected optimal performance of $k$ independent single-phase estimation with $N/k$ resources each. The mutual information in this latter case generalizes the one shown in Eq.~(\ref{2singind}) and the ultimate bound on that reads
\begin{equation}
    k\cdot I_1(N/k) \leq k(\log_2N-\log_2k).
\end{equation}
We readily find that the multiphase advantage $\Delta I_k \equiv \text{HB}(k,N) - k\cdot \text{HB}(1,N/k)$ depends on the number of simultaneously encoded phases only, namely
\begin{equation}
    \Delta I_k = k\log_2k - \log_2(k!)
\end{equation}
implying that the information gain for each phase reads $\Delta I_k/k = \log_2k - \log_2(k!)/k$, which is a positive decreasing function. Asymptotically, through Stirling's approximation, we find that the gain approaches a small constant, i.e.
\begin{equation}
    \Delta I_k/k \xrightarrow[k\gg 1]{} \,\, \log_2(e).
\end{equation}
Therefore, we can conclude that we do get an advantage from multiphase estimation, but asymptotically it is reduced to a small constant gain, as found also in Ref.~\cite{rafal2}.

\section{Conclusions} \label{7}
Exploiting the generalization of Holevo's POVM to the multiphase scenario devised in Ref.~\cite{multipe}, we developed a protocol for global multiphase estimation. The protocol generalizes the analysis in Ref.~\cite{multipe} to arbitrary cost functions by optimizing over the set of probe states. In particular, differently from Ref.~\cite{multipe}, here we considered the case of discrete estimators and took the mutual information between phases and estimators as a figure of merit, thus extending the single-phase digital estimation in Ref.~\cite{dqe}. Remarkably, the protocol performs a global estimation, therefore it can be used to test new estimation strategies in terms of the desired cost function without any a-priori notion of the phase to be estimated. \\
We showed that in the single-phase case we retrieve known optimal estimation strategies. Specifically, optimal parallel-separable and sequential (QPEA) strategies were investigated, also in the non-asymptotic case.
\\
More importantly, our protocol provides a multiphase estimation, without assuming any specific encoding strategy. We analyzed here the most basic non-trivial case, i.e. the case of two unknown phases, and pointed out the performance of generalized sequential and parallel-separable strategies in terms of mutual information.
\\
We also retrieved the ultimate bound on precision in terms of mutual information when a generic number of phases is jointly estimated given a fixed number of resources. Our bound contributes to clarify the controversy about the advantage of multiphase estimation \cite{datta,datta2,datta3,pezze,rafal2}.

\acknowledgments
This material is based upon work supported by the U.S. Department
of Energy, Office of Science, National Quantum Information Science
Research Centers, Superconducting Quantum Materials and Systems Center
(SQMS) under contract number DEAC02-07CH11359, by the EU H2020
QuantERA ERA-NET Cofund in Quantum Technologies project QuICHE and by
the National Research Centre for HPC, Big Data and Quantum Computing (ICSC). R.R is supported by the National Research Foundation, Singapore and A*STAR under its CQT Bridging Grant. We thank Francesco Albarelli for fruitful discussion.

\section{Appendix}

\subsection{Discretization} \label{appA}
\subsubsection{Conditions for the invariance of cost functions under discretization}
In parameter estimation, the mutual information is the number of bits in common between the parameter to be estimated and a suitable estimator. In many applications, the parameter is a continuous quantity but the estimator is a map from the parameter to a discrete set, which intrinsically limits the precision of the estimation. This is a realistic occurrence: achieving the required number of decimal digits for the estimation of a parameter is one of the most important issues in metrology.
\\
In this context, the discretization is determined by the number of qudits $N$ that we can exploit for our estimation. The estimator is then defined by the partition in $N$ subset of the parameter set $[0,2\pi)$, i.e. $\tilde{\phi}=2\pi m/(N+1)$ with $m\in [0,N]$.
\\
First, we need to map the continuous probability distributions $\text{Pr}(\tilde{\phi|\phi})$ into discrete quantities $p(m|\phi)$. This can be done by taking the limit
\begin{equation}
    \lim_{N\rightarrow\infty}\text{Pr}\left(\frac{2\pi\,m}{N+1} \bigg\rvert\phi\right) = \text{Pr}(\tilde{\phi}|\phi) \equiv \alpha\lim_{N\rightarrow\infty} p(m|\phi)
\end{equation}
where $\alpha$ is a normalization constant.
Indeed we also need the discrete quantities $p(m|\phi)$ to be probabilities. Then, for every fixed phase $\phi = \bar{\phi}$ we require $0\leq p(m|\bar{\phi})\leq 1 \,\forall\,\,m\in[0,N]$ and $\sum_{m=0}^{N}p(m|\bar{\phi}) = 1$. The normalization over the discrete set of the outcomes implies

\begin{equation} \label{calp2}
\begin{aligned}
    1=\int_{0}^{2\pi} \text{Pr}(\tilde{\phi}|\bar{\phi})\frac{d\tilde{\phi}}{2\pi} &= \lim_{N\rightarrow\infty} \frac{1}{N+1}\sum_{m=0}^N \text{Pr}\left(\frac{2\pi\,m}{N+1}\bigg\rvert\bar{\phi}\right) \\
    & = \lim_{N\rightarrow\infty}\frac{\alpha}{N+1}\sum_{m=0}^N p(m|\bar{\phi})
\end{aligned}
\end{equation}
implying $\alpha=N+1$ and the prescription 
\begin{equation} \label{probsymstates}
    p(m|\phi) =  \frac{1}{N+1}\text{Pr}\left(\tilde{\phi}=\frac{2\pi\,m}{N+1}\bigg\rvert\phi\right).
\end{equation}
\\
The same distribution can be derived by equivalently discretizing Holevo's POVM $d\mu(\tilde{\phi})\rightarrow \mu(m)$ through the same procedure, i.e.
\begin{equation}
\begin{aligned}
    d\mu(\tilde{\phi}) = \frac{d\tilde{\phi}}{2\pi}|e(\tilde{\phi})\rangle\langle e(\tilde{\phi})| &= \lim_{N\rightarrow\infty} \frac{1}{N+1}|e(m)\rangle\langle e(m)| \\
    &\equiv \lim_{N\rightarrow\infty} \mu(m).
\end{aligned}
\end{equation}
\\
Note that in general the distribution in Eq.~(\ref{probsymstates}) is normalized just in the limit of large $N$. However, one can check that for the distributions $p^{(1)}$ and $p^{(2)}$ we consider here $\sum_{m=0}^{N}p^{(1)}(m|\phi) = \sum_{m=0}^{N}p^{(1)}(m|\phi) \sim 1 \,\forall \,\,N$ because the width of their peaks is limited to the interval $m\in(\phi(N+1)-1,\phi(N+1)+1)$ and is the effective support of the distributions, meaning that $p^{(x)}(m\neq\phi(N+1))=o((N+1)^{-2})$ while $p^{(x)}(m=\phi(N+1))\sim 1$, with $x=1,2$.
\\
However, we still need to check how discretization affects the cost functional in Eq.~(\ref{bmd}), which shall now be expressed as
\begin{equation} \label{disc}
\begin{aligned}
    \bar{C} 
    & = \int_{0}^{2\pi}\frac{d\phi}{2\pi}\int_{0}^{2\pi}\frac{d\tilde{\phi}}{2\pi}\,\text{Pr}(\tilde{\phi}|\phi)\,c(\phi,\tilde{\phi}) \\ 
    & = \lim_{N\rightarrow\infty}\frac{1}{N+1}\int_{0}^{2\pi}\frac{d\phi}{2\pi}\sum_{m=0}^{N}\text{Pr}(m|\phi)\,c(\phi,m) \\
    &= \lim_{N\rightarrow\infty}\int_{0}^{2\pi}\frac{d\phi}{2\pi}\sum_{m=0}^N p(m|\phi)\,c(\phi,m).
\end{aligned}
\end{equation}
The form of the cost function $c(\phi,\tilde{\phi})$ is crucial to the difference between continuous and discrete case. Through the following observations, we will define the cases where the discretization works for every $N$.
\begin{proposition} \label{prop}
Consider a two-variable function $f(\phi,\tilde{\phi})\equiv \text{Pr}(\tilde{\phi}|\phi)\,c(\phi,\tilde{\phi})$, defined as the product of a conditional density and a generic function $c$. Let the two variables share a continuous domain. If
\begin{itemize}
    \item $f$ depends only on the difference between the two variables, i.e. $f(\phi,\tilde{\phi})=f(\phi-\tilde{\phi})$ \\
    \item $f$ is periodic \\ 
    \item $c(\phi,\tilde{\phi})$ does not depend on $\text{Pr}(\tilde{\phi}|\phi)$
    \item the discrete distribution implied by Eq.~(\ref{probsymstates}) is normalized $\forall N$
\end{itemize}
 then
\begin{equation}
\begin{aligned}
   &\int_{0}^{2\pi}\frac{d\tilde{\phi}}{2\pi}\int_{0}^{2\pi}\frac{d\phi}{2\pi}\,f(\phi-\tilde{\phi}) = \\
   &= \frac{1}{N+1}\int_{0}^{2\pi}\frac{d\phi}{2\pi}\,\sum_{m=0}^Nf\left(\phi-\frac{2\pi m}{N+1}\right) \quad \forall\,N.
\end{aligned}
\end{equation}
\end{proposition} 
This result in our case implies
\begin{equation}
\begin{aligned}
    \bar{C} &= \int_{0}^{2\pi}\frac{d\phi}{2\pi}\int_{0}^{2\pi}\frac{d\tilde{\phi}}{2\pi}\,\text{Pr}(\tilde{\phi}|\phi)\,c(\phi,\tilde{\phi}) \\
    &= \int_{0}^{2\pi}\frac{d\phi}{2\pi}\sum_{m=0}^N p(m|\phi)\,c(\phi,m)
\end{aligned}
\end{equation}
i.e. we can ignore the limit $N\rightarrow\infty$ in Eq.~(\ref{disc}). 
\\
\begin{proof}
Let $\tilde{\phi}$ be a continuous variable, periodic with period $2\pi$. Then
\begin{equation}
\begin{aligned} \label{intermediate}
    \int_{0}^{2\pi}\frac{d\tilde{\phi}}{2\pi}\int_{0}^{2\pi}\frac{d\phi}{2\pi}\,f(\phi-\tilde{\phi}) &= \int_{0}^{2\pi}\frac{d\gamma}{2\pi}\int_{0}^{2\pi}\frac{d\phi}{2\pi}\,f(\gamma) = \\
    &=\int_{0}^{2\pi}\frac{d\gamma}{2\pi}\, f(\gamma).
\end{aligned}
\end{equation}
On the other hand, if $\tilde{\phi}$ is discrete and, in particular, $\tilde{\phi}=2\pi m/(N+1)$, we take the sum
\begin{equation}
\begin{aligned}
    &\sum_{m=0}^N\int_{0}^{2\pi}\frac{d\phi}{2\pi}\,f\left(\phi - \frac{2\pi m}{N+1}\right) = \\
    &=\sum_{m=0}^N\int_{-\frac{2\pi m}{N+1}}^{2\pi(1-\frac{ m}{N+1})}\frac{d\gamma}{2\pi}\, f(\gamma) = \\
    &=(N+1)\int_{0}^{2\pi}\frac{d\gamma}{2\pi}\, f(\gamma) = \\
    &=(N+1) \int_{0}^{2\pi}\frac{d\tilde{\phi}}{2\pi}\int_{0}^{2\pi}\frac{d\phi}{2\pi}\,f(\phi-\tilde{\phi})
\end{aligned}
\end{equation}
where we used the periodicity of $f$ in the second equality and the result of Eq.~(\ref{intermediate}) in the third one. The third and fourth condition guarantee that the ratio over $N+1$ normalizes the resulting discrete probability according to the prescription in Eq.~(\ref{probsymstates}). 
\end{proof}
Cost functions belonging to the Holevo class, like $c(\phi,\tilde{\phi})=4\sin{(\phi-\tilde{\phi})^2/2}$, all satisfy the hypotheses of Proposition~(\ref{prop}). In such case, for any number of qubits $N$ employed for estimation, continuous and discrete estimators are equivalent to the cost function.
\subsubsection{Effects on the mutual information}
Here we are interested in the mutual information, whose related cost functional is the conditional entropy $H(\tilde{\phi}|\phi)=-\text{Pr}(\tilde{\phi}|\phi)\log \text{Pr}(\tilde{\phi}|\phi)$. Note that, as long as $\text{Pr}(\tilde{\phi}|\phi)$ is a periodic function, this functional satisfies the second of the conditions required in Proposition~(\ref{prop}).
However, the cost function related to $H(\tilde{\phi}|\phi)$ is the so-called \textit{surprise} $c(\phi,\tilde{\phi})=-\log \text{Pr}(\tilde{\phi}|\phi)$, which depends on the conditional density $\text{Pr}(\tilde{\phi}|\phi)$, against the third condition, implying that the discretized distribution defining the surprise is not normalized. Indeed, continuous and discrete estimators provide different conditional entropies. Notwithstanding, the mutual information is invariant under discretization, as we show in the following.
\begin{proposition}
Given a periodic and covariant conditional density $P(\tilde{\phi}|\phi)$, the mutual information between an estimator $\tilde{\phi}$ and a real parameter $\phi\in[0,1]$, defined as $I(\tilde{\phi}:\phi)\equiv H(\tilde{\phi})-H(\tilde{\phi}|\phi)$ is invariant under discretization of the estimator.
\end{proposition}
\begin{proof}
Let $\tilde{\phi}$ be a continuous variable. Then the conditional entropy $H(\tilde{\phi}|\phi)$ reads\begin{equation} \label{h1}
    H(\tilde{\phi}|\phi)=-\int_0^{1}d\phi\int_0^{1}d\tilde{\phi}\,\text{Pr}(\tilde{\phi}|\phi)\log \text{Pr}(\tilde{\phi}|\phi).
\end{equation}
We assume to have no prior information on the estimator. Thus the related probability distribution is uniform and the differential entropy $H(\tilde{\phi})$ depends on the integration domain. Here we have
\begin{equation}
    H(\tilde{\phi})=-\int_0^1 \text{Pr}(\tilde{\phi})\log\text{Pr}(\tilde{\phi}) = 0.
\end{equation}
Therefore, we find for the mutual information
\begin{equation} \label{mut1}
\begin{aligned}
    I(\tilde{\phi}:\phi)=-H(\tilde{\phi}|\phi) &= \int_0^{1}d\phi\int_0^{1}d\tilde{\phi}\,\text{Pr}(\tilde{\phi}|\phi)\log \text{Pr}(\tilde{\phi}|\phi) = \\
    &=\int_0^{1}d\gamma \text{Pr}(\gamma)\log \text{Pr}(\gamma)
\end{aligned}
\end{equation}
where in the last equality we exploited covariance.
Now take a discrete estimator $\tilde{\phi}=m/(N+1)$. Without assuming any convergence to Eq.~(\ref{h1}), we write the conditional entropy as
\begin{equation}
\begin{aligned}
    H(m|\phi)&=-\sum_{m=0}^N\int_0^{1}d\phi\,p(m|\phi)\log p(m|\phi) \\
    &=-\sum_{m=0}^N\int_0^{1}d\phi\,\frac{\text{Pr}(m|\phi)}{N+1}\log \frac{\text{Pr}(m|\phi)}{N+1} \\
    &= -\sum_{m=0}^N\int_{-\frac{m}{N+1}}^{1-\frac{m}{N+1}}d\gamma\,\frac{P(\gamma)}{N+1}\log \frac{P(\gamma)}{N+1} \\
    &=\log(N+1) - \int_0^1 d\gamma \text{Pr}(\gamma)\log \text{Pr}(\gamma).
\end{aligned}
\end{equation}
On the other hand, the assumption of uniform prior distribution for the estimator in the discrete case provides 
\begin{equation}
    H(m) = -\sum_{m=0}^N p(m)\log p(m) = \log (N+1)
\end{equation}
implying 
\begin{equation}
\begin{aligned}
    I(m:\phi) &= H(m) - H(m|\phi) = \\
    &=\int_0^1 d\gamma \text{Pr}(\gamma)\log \text{Pr}(\gamma) = I(\tilde{\phi}:\phi) \,\,\, \forall N
\end{aligned}
\end{equation}
which completes the proof.
\end{proof}

\subsection{Entanglement of the locally optimal input state} \label{appB}
In this appendix, we discuss the entanglement content of the states introduced in the main text. In particular, we analytically evaluate the $N$-partite entanglement among the $N$ probes of the states that are locally optimal for the mutual information, i.e. the Holland-Burnett states. We show that, as expected, the qutrit states are more entangled than the qubit ones. We use as multipartite entanglement quantifier the geometric measure of entanglement~\cite{wei2003geometric} which can be easily computed in this case due to the symmetry of the states. The geometric measure of entanglement of a pure state $|\psi\rangle$ is
\begin{equation}
    E_G(|\psi\rangle)=1-\max_{|\Phi\rangle\in \text{PRO}}|\langle\Phi|\psi\rangle|^2
\end{equation}
where we denoted with PRO the set of pure product states. The geometric measure of the state $|\Psi_0\rangle^{(1)}$ in equation~\eqref{parprobe} is zero since the state is separable. The state $|\Psi_0\rangle^{(2)}$ in Eq.~\eqref{seqprobe} is pure and symmetric and  hence the closest pure product state is symmetric~\cite{zhu2010additivity}. We have~\cite{martin2010multiqubit}
\begin{equation}
\label{qubit}
    E_G(|\Psi_0\rangle^{(2)}) \simeq 1-\frac{\sqrt{2\pi N}}{N+1} 
\end{equation}

In the qutrit case, we have have similarly that the geometric measure of the state $|\Psi_0\rangle^{(1)}$ in Eq.~\eqref{st1} is zero since the state is separable. We now calculate the geometric measure of entanglement of the state $|\Psi_0\rangle^{(2)}$ of Eq.~\eqref{entangled}. We first note that we can write the closest symmetric product state as $|\Phi\rangle = (\sqrt{p}|0\rangle+\sqrt{q}e^{i\phi}|1\rangle+\sqrt{1-p-q}e^{i\theta}|2\rangle)^{\otimes n}$. We then obtain 
\begin{align}
     &E_G(|\Psi_0\rangle^{(2)}) =  \\
     &\quad  1- \frac{1}{M}\bigg|\sum_{n_1=0}^{N}\sum_{n_2=0}^{N-n_1}\sqrt{M(N,n_1,n_2)}\sqrt{p}^{\, N-n_1-n_2}\\
     & \qquad \qquad \qquad  \times \sqrt{q}^{\,n_1} \sqrt{1-p-q}^{\, n_2} e^{-in_1\phi}e^{-in_1\theta}\bigg|^2
\end{align}
It is easy to see that th previous expression 
is minimised for $\phi=\theta=0$ and $p=q=1/3$. We can approximate the sum by an integral and the  multinomial distribution by a 2-D Gaussian with mean $\vec{\mu}=N(1/3,1/3)$ and variance $\Sigma = N((2/9,-1/9),(-1/9,2/9))$. Straightforwardly, we get
\begin{equation}
\label{qutrit}
     E_G(|\Psi_0\rangle^{(2)}) \simeq 1 - \frac{8\pi N}{3\sqrt{3} M}
\end{equation}
The above result implies that the qutrit states are more entangled than the qubit ones.
Note that the relative entropy of entanglement is divergent in both cases since $E(\rho) \geq -\log{(1-E_G(\rho))}$.

\begin{figure}
    \centering
    \includegraphics[scale=.45]{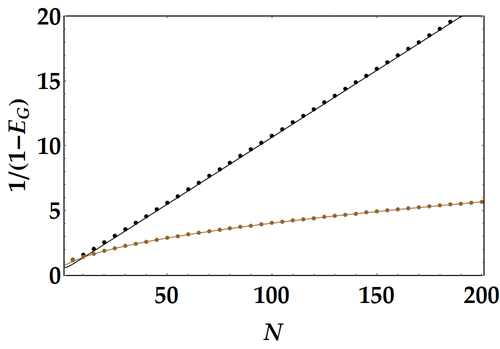}
    \caption{Entanglement of the state $|\Psi_0\rangle^{(2)}$ in the qubit (brown) and qutrit (black) cases as a function of $N$. The solid black and the brown lines are the asymptotic behaviors in Eqs.~\eqref{qubit} and~\eqref{qutrit}, respectively, while the dots displays the corresponding the exact values. For the qubit case the inverse of the fidelity of separability (i.e. $1/(1-E_G)$) scales as $\sqrt{N}$. In the qutrit case it is linear in $N$.}
    \label{prob}
\end{figure}

\subsection{Heisenberg bound on the mutual information for multiphase estimation} \label{appC}
Here we provide the proof of Proposition~\ref{prop2}. We follow the approach given in \cite{supp}. 
\begin{proof}
Be $\pmb{\phi}=(\phi_1,\phi_2,...,\phi_k)$ the vector of phases we wish to estimate. Be the unitary representation $U_{\pmb{\phi}}\equiv \exp{(i\sum_{j=1}^kH_j\phi_j)}$, with $\{H_j\}_{j=1}^k$ as a set of commuting generators $H_j=|j\rangle\langle j|$. Take a generic probe state $\rho_0 \in \mathcal{S}(\mathcal{H}_{||})$ where $\mathcal{S(\cdot)}$ is a set of states, here defined on the Hilbert space $\mathcal{H}_{||}$ spanned by the non-degenerate vectors $|\pmb{n}\rangle\equiv \{|n_j\rangle_{||}\}_{j=0}^{k}$ with the constraint on the sum of the corresponding eigenvalues given by $\sum_{j=0}^{k}n_j = N$, i.e. $n_0 = N - \sum_{j=1}^{k}n_j$. Then $\rho_0$ can be expressed as
\begin{equation}
    \rho_0 = \sum_{\pmb{n},\pmb{m}}\lambda_{\pmb{n},\pmb{m}}|\pmb{n}\rangle\langle\pmb{m}|
\end{equation}
with $\sum_{\pmb{n}}\equiv \sum_{n_1=0}^{N}\sum_{n_2=0}^{N-n_1}...\sum_{n_{k}=0}^{N-\sum_{j=1}^{k-1}n_j}$ and $\sum_{\pmb{n}}\lambda_{\pmb{n}, \pmb{n}}=1$.
Note that we set the number of vectors $|n_j\rangle_{||}$ to $k+1$ in order to exploit the whole basis of the non-degenerate subspace, except for a reference state, to encode the phases to be estimated.
\\
The application of the unitary on the probe state yields 
\begin{equation}
    \rho_{\pmb{\phi}} = U_{\pmb{\phi}}^{\otimes N}\rho_0(U_{\pmb{\phi}}^{\dagger})^{\otimes N}. 
\end{equation}
Be $p_{\pmb{\phi}}=p(\phi_1,\phi_2,...,\phi_k)$ the corresponding joint probability distribution. Finally, let $S(\rho)$ be the Von Neumann entropy related to state $\rho$. Then the maximum information on $\pmb{\phi}$ extractable through a corresponding set of estimators $\pmb{\tilde{\phi}}$ from $\rho_{\pmb{\phi}}$ is given by the Holevo bound, i.e.
\begin{equation}
    I(\pmb{\tilde{\phi}}:\pmb{\phi}) \leq S\left(\int\,d\pmb{\phi}\,p_{\pmb{\phi}}\rho_{\pmb{\phi}}\right) - \int\,d\pmb{\phi}\,p_{\pmb{\phi}}S(\rho_{\pmb{\phi}}).
\end{equation}
We remark that the bound is maximized for pure states, since, for pure $\rho$, we have $S(\rho)=0$. Therefore, henceforth we take the probe in a pure state.
In this case, the Holevo bound reads $I(\pmb{\tilde{\phi}}:\pmb{\phi}) \leq S(\int\,d\pmb{\phi}\,p_{\pmb{\phi}}\rho_{\pmb{\phi}})$. Note that the mutual information $I(\pmb{\tilde{\phi}}:\pmb{\phi})$ is the same one that in the main text we named $I_k(N)$ to emphasize the main parameters, i.e. the number of phases to be estimated $k$ and the number of resources $N$.
\\
Now consider the projectors over the eigenbases of the generators defined as $P_{\pmb{n}}^{(k)}\equiv |\pmb{n}\rangle\langle \pmb{n}|$. Since the sum over the eigenvalues $\{n_j\}_{j=0}^k$ is constrained to be equal to the number of resources $N$, the number of projectors $P_{\pmb{n}}^{(k)}$ is 
\begin{equation} \label{m}
    M \equiv \sum_{n_1=0}^{N}\sum_{n_2=0}^{N-n_1}...\sum_{n_k=0}^{N-\sum_{j=1}^{k-1}n_j}1 = \binom{N+k}{N}
\end{equation}
which is the dimension of $\mathcal{H}_{||}$ \cite{werner}.\\
The entropy bounding the mutual information can be bounded as
\begin{equation}
    \begin{aligned} 
       S\left(\int\,d\pmb{\phi}\,p_{\pmb{\phi}}\rho_{\pmb{\phi}}\right) &= S\left(\int\,d\pmb{\phi}\,p_{\pmb{\phi}}U_{\pmb{\phi}}^{\otimes N}\rho_0(U_{\pmb{\phi}}^{\dagger})^{\otimes N}\right) \\
       &\leq S\left(\int\,d\pmb{\phi}\,p_{\pmb{\phi}}\sum_{\pmb{n}}P^{(k)}_{\pmb{n}}U_{\pmb{\phi}}^{\otimes N}\rho_0(U_{\pmb{\phi}}^{\dagger})^{\otimes N}P^{(k)}_{\pmb{n}}\right) \\
       &= S\left(\int\,d\pmb{\phi}\,p_{\pmb{\phi}}\sum_{\pmb{n}}P^{(k)}_{\pmb{n}}\rho_0P^{k}_{\pmb{n}}\right) \\
       &= S(\sum_{\pmb{n}}P^{(k)}_{\pmb{n}}\rho_0P^{(k)}_{\pmb{n}})  \\
       &= S(\sum_{\pmb{n}}\lambda_{\pmb{n},\pmb{n}}|\pmb{n}\rangle\langle\pmb{n}|) \\
       &= H(\lambda_{\pmb{n},\pmb{n}}) + \sum_{\pmb{n}}\lambda_{\pmb{n},\pmb{n}}S(|\pmb{n}\rangle\langle\pmb{n}|) \\
       &= H(\lambda_{\pmb{n},\pmb{n}}) \leq \log_2(M) \label{holeheise}
    \end{aligned}
\end{equation}
where, in the second line we used the data-processing inequality for projectors and in the fourth line the normalization of the joint probability distribution $\int\,d\pmb{\phi}\,p_{\pmb{\phi}} = 1$. The quantity $H(\lambda)$ appearing in the last two lines is the Shannon entropy $H(\lambda)\equiv -\sum_k\lambda_k\log_2\lambda_k$.
\\
Note that the number of projectors included in the second line is strictly connected to the number of encoded phases $k$, accordingly with the definition of $M$ in Eq.~(\ref{m}) which eventually provides the bound
\begin{equation}
    I(\pmb{\tilde{\phi}}:\pmb{\phi}) \leq \log_2\binom{N+k}{N} \label{final}.
\end{equation}
and completes the proof.

\end{proof}

\bibliography{paper}

\end{document}